# Microscopic mechanism of low thermal conductivity in lead-telluride

Takuma Shiga[1,2], Junichiro Shiomi[1,2,†], Jie Ma[3], Olivier Delaire[3], Tomasz Radzynski[4], Andrzej Lusakowski[4], Keivan Esfarjani[5], Gang Chen[5]

[1]*Department of Mechanical Engineering, The University of Tokyo, Hongo Bunkyo-ku, Tokyo, 113-8656, Japan*

[2]*Japan Science and Technology Agency, PRESTO, 4-1-8 Honcho, Kawaguchi, Saitama, 332-0012, Japan*

[3]*Neutron Scattering Science Division, Oak Ridge National Laboratory, 1 Bethel Valley Road, Oak Ridge, Tennessee 37831, USA*

[4]*Institute of Physics, Polish Academy of Sciences, Warsaw, Poland*

[5]*Department of Mechanical Engineering, Massachusetts Institute of Technology, 77 Massachusetts Avenue, Cambridge, MA 02139, USA*

**Abstract**

The microscopic physics behind low lattice thermal conductivity of single crystal rocksalt lead telluride (PbTe) is investigated. Mode-dependent phonon (normal and umklapp) scattering rates and their impact on thermal conductivity were quantified by the first-principles-based anharmonic lattice dynamics calculations that accurately reproduce thermal conductivity in a wide temperature range. The low thermal conductivity of PbTe is attributed to the scattering of longitudinal acoustic phonons by transverse optical phonons with large anharmonicity, and small group velocity of the soft transverse acoustic phonons. This results in enhancing the relative contribution of optical phonons, which are usually minor heat carrier in bulk materials.

## I. INTRODUCTION

The energy conversion efficiency of semiconducting thermoelectric materials is characterized by the dimensionless figure of merit $ZT=S^2\sigma_{el}T/(\kappa_{el}+\kappa_{lat})$, where $S$, $\sigma_{el}$, $\kappa_{el}$, $\kappa_{lat}$, and $T$ are the Seebeck coefficient, electrical conductivity, electron thermal conductivity, lattice thermal conductivity, and temperature, respectively[1]. Strategies to suppress $\kappa_{lat}$ without critically deteriorating the other properties have been shown to be effective to enhance $ZT$. This includes alloying species with mass mismatch to enhance mass-difference phonon scattering[2], filling with rattling ions to scatter phonons in cage compounds[3,4], or embedding layered structures to reduce the phonon group velocity and/or enhance intrinsic phonon scattering[5]. These approaches are complementary to recently-developed nanostructuring techniques[6,7], to improve the starting-material properties. While these intuitive approaches rely on the inhomogeneity or complexity of crystals, there are bulk materials with simple structures that show low thermal conductivity. The rocksalt lead telluride (PbTe) is a prime example among thermoelectric materials that has been long known to exhibit low thermal conductivity (~2 Wm$^{-1}$K$^{-1}$ at 300 K) despite its high symmetry and coordination.

PbTe is one of the most promising thermoelectric materials for applications in the intermediate temperature range (400-800 K) due to its low $\kappa_{lat}$, as well as high $S$ and $\sigma_{el}$ when appropriately doped[8-10]. The counter-intuitively low thermal conductivity of PbTe has been discussed based on phonon dispersion relations either calculated from first-principles[11,12] or measured by inelastic neutron scattering (INS)[13]. The first principles calculations have shown that the energy of the transverse optical (TO) phonons near the zone center ($\Gamma$ point) is highly sensitive to changes in volume[11,12], and this led researchers to conjecture a strong coupling between TO and longitudinal acoustic (LA) phonons since they are compressive waves. More recently, INS measurements[13] have probed aspects, such as the "avoided-crossing" of LA and TO branches and the complex peak broadening in INS intensity of TO modes near Γ point, which, through momentum and energy conservation arguments and mode coupling theory, are suggested to be due to strongly anharmonic LA-TO coupling.

While the idea of strong LA-TO scattering may be plausible, the arguments so far remain phenomenological without quantitative knowledge of the actual scattering rates of the processes. In addition, it is not clear, how strong the LA-TO scattering is compared with other scattering processes and how much it influences the overall thermal conductivity. In this paper, we answer these questions by calculating the scattering rates of all the three-phonon processes occurring in PbTe using the framework of first-principles-based anharmonic lattice dynamics (ALD), which has been applied successfully to simple monoatomic crystals[14-16] and relatively complex compounds[17].

## II. METHODOLOGY

To examine the anharmonic phonon dynamics, it is important to obtain accurate anharmonic interatomic force constants (IFCs), which are here defined as the Taylor expansion coefficients of the force, with respect to atomic displacements from equilibrium configurations,

$$F_{i,\alpha} = -\Pi_{i,\alpha} - \sum_{j,\beta} \Phi_{ij}^{\alpha\beta} u_{j,\beta} - \frac{1}{2}\sum_{jk,\beta\gamma} \Psi_{ijk}^{\alpha\beta\gamma} u_{j,\beta} u_{k,\gamma}, \qquad (1)$$

where $\Phi_{ij}^{\alpha\beta}$ and $\Psi_{ijk}^{\alpha\beta\gamma}$ are harmonic and cubic anharmonic IFCs, respectively. The indices $i, j$, and $k$ are the atom indices, and $\alpha, \beta$, and $\gamma$ denote the Cartesian components. In this study, the IFCs of a rock-salt PbTe were computed following the *real-space*

*displacement* method[18], with the Hellman-Feynman forces calculated by density functional theory (DFT) in a 2×2×2 conventional supercell with 64 atoms.

In the *real-space displacement* method, sets of force-displacement data are obtained by systematically displacing one or two atoms at a time about their equilibrium position along Cartesian coordinates. Here, the atoms were displaced by ± 0.01 Å and ± 0.02 Å. Then Eq. (1) can be fitted to the force-displacement data, together with the symmetry properties, translational/rotational invariance conditions, by using a singular value decomposition algorithm to obtain harmonic and cubic IFCs[18]. The ranges of IFCs (number of neighboring shells that interact with a given atom) were chosen to be respectively six and one for harmonic and cubic IFCs, to minimize the fitting residual, while keeping the computation affordable. While the method could in principle work with any number of force-displacement data due to its fitting nature, in the current work, displacements were performed for all the irreducible degrees of freedom by taking the symmetry into account.

We have used the *Quantum Espresso* package[19] for the DFT calculations. The norm conserving pseudo potential was designed to incorporate relativistic effects. We find that incorporation of the spin-orbit coupling is important to assure the level of accuracy in the calculated thermal conductivity presented later. A 5×5×5 Monkhorst-Pack k-point mesh[20] was used to sample electronic states in the first Brillouin zone, and an energy cutoff of 50 Ryd (~650 eV) was used for the plane-wave expansion. The lattice constant was set to 6.548 Å, which was found to minimize the ground state total energy. The PBE functional was used to describe the exchange-correlation effects.

Using the harmonic IFCs, the phonon dispersion relations can be readily calculated through the dynamical matrices[21]. Note that, since PbTe is a polar crystal, the effect of ionic charges needs to be considered. This was modeled, for simplicity, by adding the non-analytical term[22] to the dynamical matrix with the Born effective charges and dielectric constant $Z^*_{Pb} = +5.87$, $Z^*_{Te} = -5.87$, and $\varepsilon^\infty = 38.48$, which were calculated through macroscopic electric polarization using the density functional perturbation theory[23].

## III. RESULTS AND DISCUSSION

The phonon dispersion relations of PbTe are shown in Fig. 1(a). The calculations reproduce the overall features of the phonon dispersion relations obtained by the INS experiments[24] with particularly good agreement for the acoustic phonons, which are important as they are the main heat carriers. One noticeable feature is that the acoustic branches, particularly the TA branches, are very soft. For instance, the group velocity of the long wave TA modes along the [100] direction is 1185 m/s, comparable to the group velocity of van der Waals systems, such as solid argon[25]. Such slow propagation speed of active carriers is expected to limit the contribution of TA phonons to the overall thermal conductivity.

It can be noticed that the discrepancy between the calculated and experimental dispersion relations is larger for optical phonons, particularly for the LO branch, which splits from the TO branch at the Γ point due to the ionic charge effect. The discrepancy at the Γ point is due to the difference in the Born effective charges and dielectric constants, which in this work were calculated for ground states, and the discrepancy in the line shapes of the phonon dispersion relations away from Γ is due to the excess simplification of the non-analytical term[22]. Therefore, the sensitivity of the results to the discrepancy was checked by tuning the Born effective charges and dielectric constants to match the LO frequency at Γ with that of the experiment, and the changes in the calculated acoustic phonon scattering rates and $\kappa_{lat}$ were confirmed be less than 1%.

The cubic IFCs enable us to compute the mode Grüneisen parameters, which is a useful measure of the anharmonicity of the crystal. As shown in Fig. 1(b), the mode Grüneisen parameters of TO phonons exhibit a diverging trend upon approaching the Γ point. This has been discussed to cause significant softening of near-Γ TO modes[11, 12]. The result indicates that dynamics of TO phonons near Γ point is extremely anharmonic, which is expected to enhance their scattering rates with the acoustic phonons and thus affect the thermal conductivity.

With the eigenstates and cubic Hamiltonian, the self-energy $\Sigma_{\mathbf{q}s}$ of phonons can be calculated by anharmonic lattice dynamics (ALD) approach[26], where $\mathbf{q}$ and $s$ denote the wavevector and phonon branch, respectively. Here, we only take into account the three phonon (normal and umklapp) scattering processes, which are the dominant source of intrinsic thermal resistance[21]. The calculation is based on Fermi's golden rule of the phonon scattering processes that satisfy the energy and momentum conservation, with phonon generation and annihilation operators derived using the cubic IFCs. The calculations were done for $N_\mathbf{q} \times N_\mathbf{q} \times N_\mathbf{q}$ uniform reciprocal meshes. The imaginary part of the self-energy gives the phonon scattering rate $w_{\mathbf{q}s}$, which is inversely proportional to the phonon relaxation time, $\tau_{\mathbf{q}s}$ ($w_{\mathbf{q}s}=2\text{Im}(\Sigma_{\mathbf{q}s})=1/\tau_{\mathbf{q}s}$). It was confirmed that the calculated phonon scattering rates were well converged with $N_\mathbf{q}=20$.

Figure 2(a) shows the frequency dependence of phonon relaxation time at 300 K calculated by ALD with $N_\mathbf{q}=20$. The phonon relaxation times of acoustic branches exhibit inverse quadratic dependence on frequency ($\tau \propto \nu^2$, where $\nu$ is frequency) in the low frequency region, in agreement with Klemens' prediction[27]. In this low frequency region, where the active heat carriers reside, the phonon relaxation times of TA phonons are longer than those of LA phonons. We can further separate the scattering rates into those of normal and umklapp processes as shown in Figs. 2(b) and 2(c). The results clearly identify the scalings for normal process, $\tau_{\text{normal}} \propto \nu^{-2}$, and umklapp process, $\tau_{\text{umklapp}} \propto \nu^{-3}$ which have been derived by Esfarjani et al.[16] as leading order relations in the low frequency limit. The results show that the scattering rate of normal processes is larger in the low frequency region, and this, through the Matthiessen's rule[21], gives rise to the inverse quadratic frequency-dependence of the overall phonon relaxation time.

Let us now look into the wavenumber dependence of the scattering rate $w_{\mathbf{q}s}$ along a representative ($\Gamma$-X) symmetry line. Figures 3(a) and 3(b) show $w_{\mathbf{q}s}$ of acoustic and optical phonons, respectively, at 300 K calculated with $N_\mathbf{q}$=20. The results show that $w_{\mathbf{q}s}$ of LA phonon is significantly larger than that of TA phonon for a given wavenumber along $\Gamma$-X, which is consistent with the overall feature of phonon relaxation time shown in Fig. 2. The scattering rates of the LA phonons along $\Gamma$-X are now compared with those extracted from the INS experiment[13]. In the experiment, $w_{\mathbf{q}s}$ is obtained by fitting the line shape with a Gaussian function and extracting the linewidth, corrected for the instrument **q**-(wavevector) and E-resolution (energy), as well as the slope of the dispersion. The full-width half maximum value then corresponds to the scattering rate. As shown in Fig. 3(a), the results of the calculations and experiments exhibit reasonable agreement, particularly for the intermediate wavenumbers. The calculation gives smaller $w_{\mathbf{q}s}$ than the experiments for smaller wavenumbers presumably due to the absence of the avoided-crossing in the calculations. The results of optical phonon scattering rates also show high polarization dependence as shown in Fig. 3(b). A noticeable feature is that $w_{\mathbf{q}s}$ of TO phonon increases as **q** approaches the Γ point, which is consistent with the above mentioned divergence of mode-Grüneisen parameter.

Using the obtained phonon relaxation times, $\kappa_{\text{lat}}$ was calculated based on the relaxation time approximation of the phonon gas[26],

$$\kappa_{\text{lat}}^{N_{\mathbf{q}}} = \frac{1}{3}\sum_{\mathbf{q}s} c_{\mathbf{q}s} \mathbf{v}_{g,\mathbf{q}s}^2 \tau_{\mathbf{q}s} = \frac{1}{3}\sum_{\mathbf{q}s} c_{\mathbf{q}s} v_{g,\mathbf{q}s} \Lambda_{\mathbf{q}s}, \qquad (2)$$

where $c_{\mathbf{q}s}$, $\mathbf{v}_{g,\mathbf{q}s}$, and $\Lambda_{\mathbf{q}s}$ are the specific heat, group velocity vector, and mean free path (MFP) of phonon $\mathbf{q}s$, respectively. Since the calculation was done for discretized mesh points, one needs to account for the size effect due to the missing contribution of long wavelength acoustic phonons[16, 17]. For this, as shown in Fig. 4(a), $\kappa_{lat}^{N_{\mathbf{q}}}$ was calculated for several values of $N_{\mathbf{q}}$, and $\kappa_{\text{lat}}$ (bulk lattice thermal conductivity) was obtained by extrapolating the data using the scaling of the size effect, $\kappa_{lat}^{N_{\mathbf{q}}} \propto N_{\mathbf{q}}^{-1}$[16, 17]. From the extrapolation, it can also be noted that the discrete-mesh calculation with $N_{\mathbf{q}}$=20 recovers nearly all the phonons that contribute to $\kappa_{\text{lat}}$. As a result, the temperature dependence of the extrapolated $\kappa_{\text{lat}}$ is in excellent agreement with the two experimental results[28, 29] in a wide temperature range (100-900 K). The quantitative agreement ensures the accuracy of the current first-principles-based ALD methodology. The agreement in the high temperature region assures the validity of the approximation to incorporate IFCs up to the cubic terms in Eq. (1).

Having the microscopic information of phonon transport, we are able to break down the thermal conductivity into contributions from different phonon branches. This was done in terms of cumulative thermal conductivity, $\kappa_c$, which is defined as a summation of thermal conductivity up to MFP of $\Lambda_0$[30],

$$\kappa_c(\Lambda_0) = \frac{1}{3}\sum_{\mathbf{q}s}^{\Lambda_{\mathbf{q}s}<\Lambda_0} c(\omega_{\mathbf{q}s})v_{\mathbf{q}s}\Lambda_{\mathbf{q}s} . \qquad (3)$$

Figure 4(b) shows the $\kappa_c$ at 300 K calculated by ALD with $N_\mathbf{q}$=20. The range of LA phonon MFPs with noticeable contribution to $\kappa_{lat}$ is limited (up to about 10 nm), even similar to those of optical modes. This is much smaller than that of TA phonons (exceeding 100 nm), reflecting the difference in the scattering rates observed in Figs. 2 and 3(a). On the other hand, since the group velocity of long-wave TA phonons are much smaller than that of LA phonons, the average thermal conductivity contribution per branch ends up being similar for LA and TA modes.

To understand the cause of this unusually large LA phonon scattering, we have identified the phonon modes that are involved in the scattering of LA phonons at representative wavenumbers ($q$=0.1 0.2, and 0.5 normalized by the wavenumber at X point) along Γ-X line. The results are summarized in Table 1 in terms of the partial sum of scattering rates, calculated for scattering processes (1) with only acoustic phonons, (2) involving at least one TO phonon, and (3) involving at least one LO phonon. The fraction with respect to $w_{q\text{LA}}$ is also denoted in parentheses. The results show that the major contribution to $w_{q\text{LA}}$ comes from the scattering processes involving TO phonons, particularly for small $q$.

**IV. CONCLUSIONS**

We have performed first-principles-based ALD calculations to investigate phonon transport in rocksalt PbTe. The real-space displacement approach gives us inexpensive access to accurate harmonic and cubic IFCs. Mode-dependent phonon relaxation times were quantified in the entire first Brillouin-zone, and the inverse quadratic frequency dependence was identified to originate from the normal process. The calculated wavevector-dependent scattering rates show reasonable agreement with those measured by the INS measurements. As a result, the lattice thermal conductivity computed under the relaxation time approximation shows excellent agreement with the experimental results. Finally, by quantifying the mode dependent thermal conductivity and the scattering processes, the low thermal conductivity was attributed equally to the strongly anharmonic LA-TO scattering, and the small group velocity of the soft TA phonons.

**ACKNOWLEDGEMENTS**

This work is partially supported by Japan Science and Technology Agency, PRESTO, the Global COE Program "Global Center of Excellence for Mechanical System Innovation" (T.S. and J.S.), and the "Solid State Solar-Thermal Energy Conversion Venter ($S^3$TEC)", and Energy Frontier Research Center funded by the US Department of Energy, Office of Science, Office of Basic Energy Sciences under Award Number: DE-SC0001299 (O.D., J.M., K.E. and G.C).

† shiomi@photon.t.u-tokyo.ac.jp
1  H. J. Goldsmid, *Introduction to Thermoelectricity* (Springer, 2009).
2  A. Majumdar, Science **303**, 777 (2004).
3  D. J. Braun and W. Jeitschko, J. Less-Common Met. **72**, 147 (1980).
4  G. A. Slack and V. G. Tsoukala, J. Appl. Phys. **76**, 1665 (1994).
5  K. Koumoto, I. Terasaki, T. Kajitani, M. Ohtaki, and R. Funahashi, in *Thermoelectrics Handbook: Macro to Nano*, edited by D. Rowe (CRC Press, New York, 2005), p. 1.
6  B. Poudel, Q. Hao, Y. Ma, Y. Lan, A. Minnich, B. Yu, X. Yan, D. Wang, A. Muto, D. Vashaee, X. Chen, J. Liu, M. S. Dresselhaus, G. Chen, and Z. Ren, Science **8**, 4670 (2008).
7  K. Biswas, J. He, Q. Zhang, GuoyuWang, C. Uher, V. P. Dravid, and M. G.

| | Kanatzidis, Nat. Chem. **16**, 160 (2011). |
|---|---|
| 8 | M. G. Kanatzidis, Chemistry of Materials **22**, 648 (2009). |
| 9 | J. P. Heremans, V. Jovovic, E. S. Toberer, A. Saramat, K. Kurosaki, A. Charoenphakdee, S. Yamanaka, and G. J. Snyder, Science **321**, 554 (2008). |
| 10 | Y. Pei, X. Shi, A. LaLonde, H. Wang, L. Chen, and G. J. Snyder, Nature **473**, 66 (2011). |
| 11 | Y. Zhang, X. Ke, C. Chen, J. Yang, and P. R. C. Kent, Physical Review B **80**, 024304 (2009). |
| 12 | J. An, A. Subedi, and D. J. Singh, Solid State Communications **148**, 417 (2008). |
| 13 | O. Delaire, J. Ma, K. Marty, A. F. May, M. A. McGuire, M. H. Du, D. J. Singh, A. Podlesnyak, G. Ehlers, M. D. Lumsden, and B. C. Sales, Nat. Mater. **10**, 614 (2011). |
| 14 | D. A. Broido, M. Malorny, G. Birner, N.Mingo, and D. A. Stewart, Applied Physics Letters **91**, 231922 (2007). |
| 15 | A. Ward, D. A. Broido, D. A. Stewart, and G. Deinzer, Physical Review B **80**, 125203 (2009). |
| 16 | K. Esfarjani, G. Chen, and H. T. Stokes, Physical Review B **84**, 085204 (2011). |
| 17 | J. Shiomi, K. Esfarjani, and G. Chen, Physical Review B **84**, 104302 (2011). |
| 18 | K. Esfarjani and H. T. Stokes, Physical Review B **77**, 144112 (2008). |
| 19 | G. Paolo, B. Stefano, B. Nicola, C. Matteo, C. Roberto, C. Carlo, C. Davide, L. C. Guido, C. Matteo, D. Ismaila, C. Andrea Dal, G. Stefano de, F. Stefano, F. Guido, G. Ralph, G. Uwe, G. Christos, K. Anton, L. Michele, M.-S. Layla, M. Nicola, M. Francesco, M. Riccardo, P. Stefano, P. Alfredo, P. Lorenzo, S. Carlo, S. Sandro, S. Gabriele, P. S. Ari, S. Alexander, U. Paolo, and M. W. Renata, Journal of Physics: Condensed Matter **21**, 395502 (2009). |
| 20 | H. J. Monkhorst and J. D. Pack, Physical Review B **13**, 5188 (1976). |
| 21 | N. Ashcroft and D. Mermin, *Solid state physics* (Thomson Learning, 1976). |
| 22 | K. Parlinski, Z.-Q. Li, and Y. Kawazoe, Physical Review Letters **78**, 4063 (1997). |
| 23 | S. Baroni, S. de Gironcoli, A. Dal Corso, and P. Giannozzi, Reviews of Modern Physics **73**, 515 (2001). |
| 24 | W. Cochran, R. A. Cowley, G. Dolling, and M. M. Elcombe, Proceedings of the Royal Society of London. Series A. Mathematical and Physical Sciences **293**, 433 (1966). |
| 25 | G. J. Keeler and D. N. Batchelder, J. Phys. C: Solid State Phys. **3**, 510 (1970). |
| 26 | G. P. Srivastava, *The Physics of Phonons* (Taylor & Francis, 1990). |

27  P. G. Klemens, Proceedings of the Royal Society of London. Series A. Mathematical and Physical Sciences **208**, 108 (1951).
28  J. R. Sootsman, R. J. Pcionek, H. Kong, C. Uher, and M. G. Kanatzidis, Chemistry of Materials **18**, 4993 (2006).
29  I. U. I. Ravich, B. A. Efimova, and I. A. Smirnov, *Semiconducting lead chalcogenides* (Plenum Press, 1970).
30  C. Dames and G. Chen, in *Thermoelectrics Handbook: Macro to Nano*, edited by D. Rowe (CRC Press, 2005).

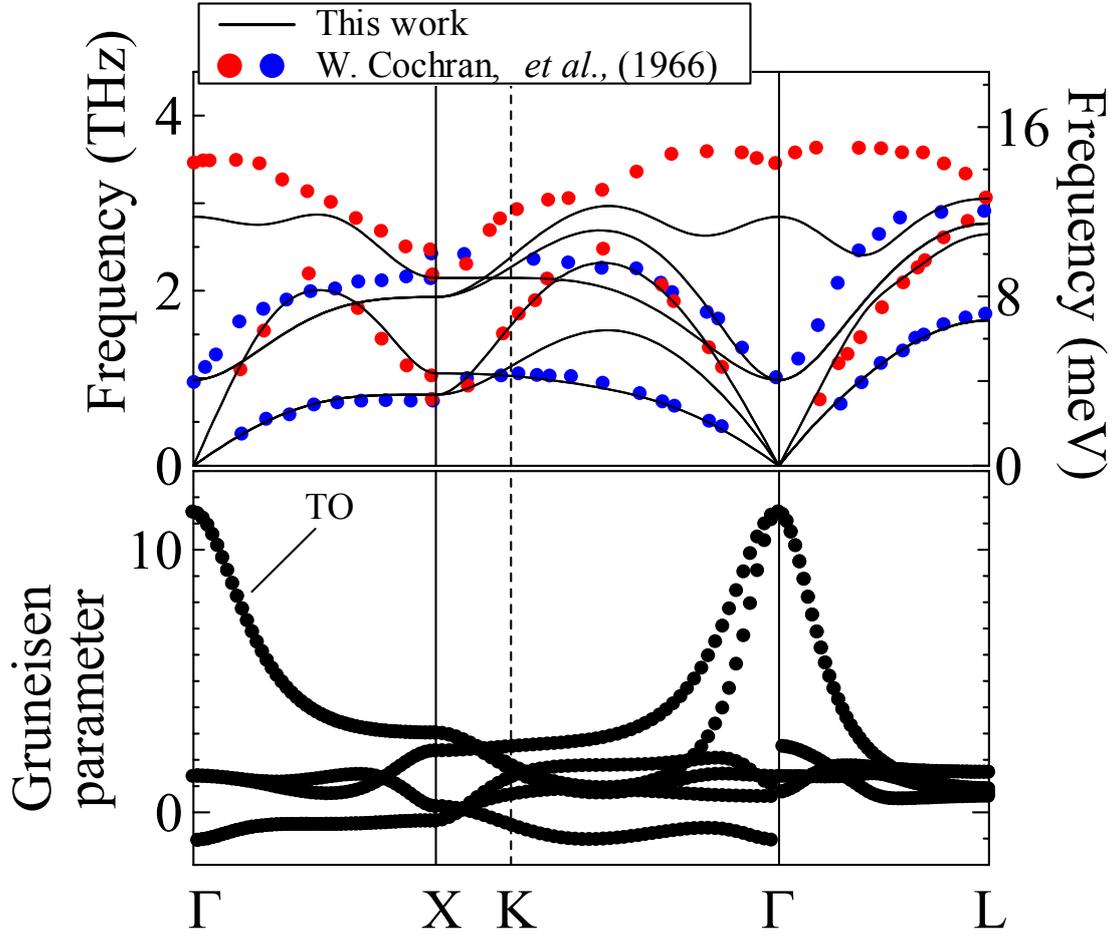

**FIG. 1.** (Color Online) (a) Phonon dispersion relations of PbTe along Γ-X-K-Γ-L symmetry lines. Red and blue filled circles indicate transverse and longitudinal modes, respectively, from inelastic neutron scattering (INS) experiments at 300 K[24]. (b) The mode-Grüneisen parameters.

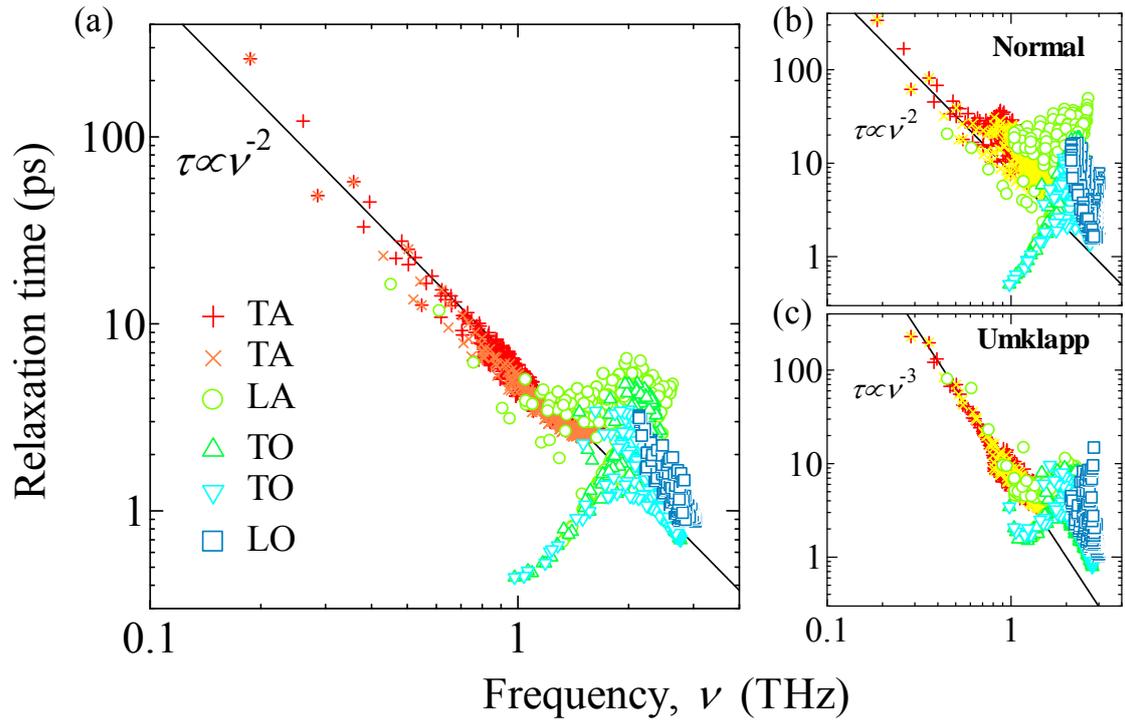

**FIG. 2.** (Color Online) (a) Frequency-dependent phonon relaxation times of PbTe at 300 K by anharmonic lattice dynamics (ALD) calculations with $N_\mathbf{q}=20$. Frequency-dependent phonon relaxation times of (b) normal and (c) umklapp processes, respectively. The solid lines denote (a) $\tau = 6\times10^{12}\nu^{-2}$, (b) $\tau = 8.0\times10^{12}\nu^{-2}$, and (c) $\tau = 8.0\times10^{24}\nu^{-3}$.

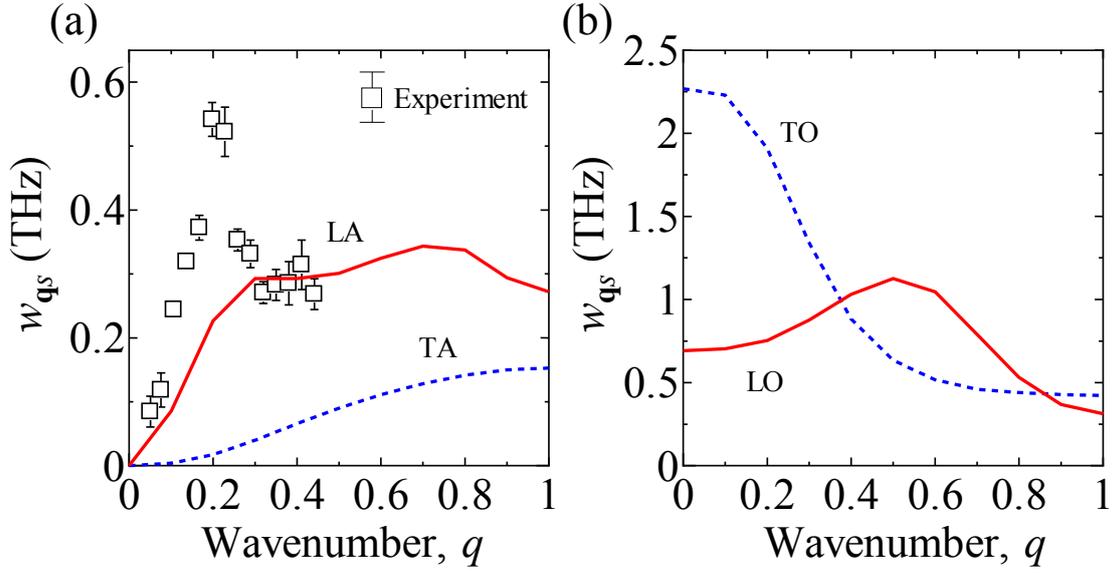

**FIG. 3.** (Color Online) The scattering rate $w_{\mathbf{q}s}$ of (a) acoustic and (b) optical phonons along Γ-X line at 300 K calculated by ALD with $N_q=20$. The inset figure in (a) shows $w_{\mathbf{q}s}$ of LA phonons at 300 K from INS experiments[13]. The wavenumber $q$ is normalized by the wavenumber of X point.

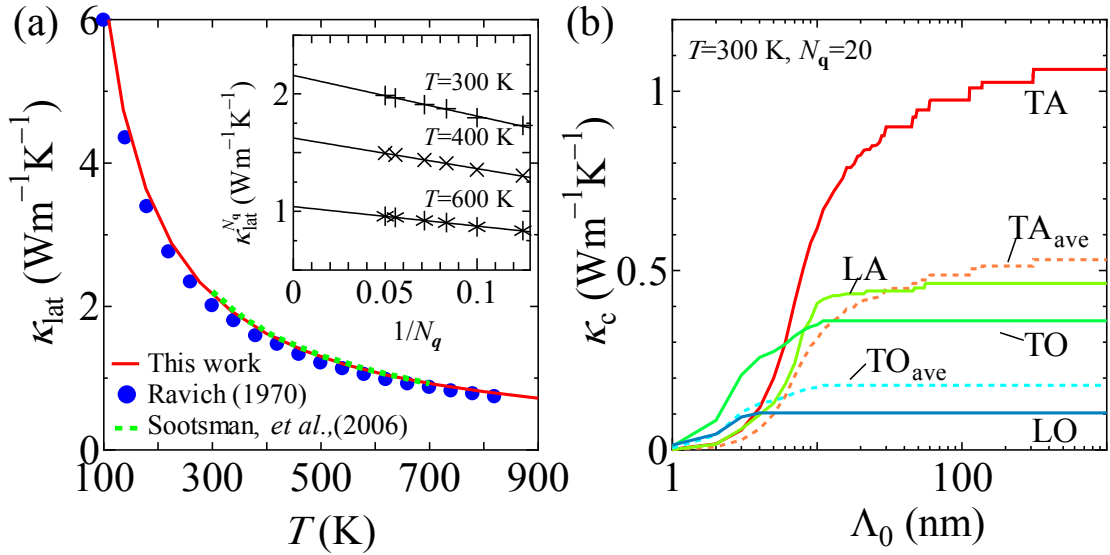

**FIG. 4.** (Color Online) (a) Comparison between the temperature-dependent bulk thermal conductivity $\kappa_{lat}$ of PbTe obtained by the ALD calculations (square) and the experiments (circles[28] and dashed line[29]). The inset figure shows the extrapolation of $\kappa_{lat}^{N_q}$ performed to obtain the bulk thermal conductivity $\kappa_{lat}$ at different temperatures. (b) Cumulative thermal conductivity $\kappa_c$ as a function of mean free path (MFP) at 300 K calculated by ALD with $N_q=20$. TA$_{ave}$ denotes the averaged values of the two TA branches.

| Partial scattering rates of LA phonon along Γ-X (THz) | | | |
|---|---|---|---|
| Scattering processes | $q$=0.1 | $q$=0.2 | $q$=0.5 |
| (1) Only Acoustic | 0.025  (29.6 %) | 0.045  (19.8 %) | 0.087  (28.9 %) |
| (2) TO involved | 0.053  (61.5 %) | 0.144  (63.4 %) | 0.122  (40.3 %) |
| (3) LO involved | 0.014  (16.2 %) | 0.056  (24.8 %) | 0.127  (42.1 %) |

**Table 1.** The impact of various scattering processes on $w_{q\text{LA}}$, for $q$ =0.1, 0.2, and 0.5 at 300 K. Partial scattering rates (THz) were calculated for (1) scattering with only acoustic phonons (2) scattering involving at least one TO phonon, and (3) scattering involving at least one LO phonon. The fractions in the parentheses are normalized by $w_{q\text{LA}}$ and thus the overall sum exceeds 100 % since the categories are not mutually exclusive, but remains below ~110 %.